\documentclass[superscriptaddress, reprint, amsmath, amssymb, aps, pra, floatfix]{revtex4-2}

\usepackage[utf8]{inputenc}
\usepackage{graphicx}% Include figure files
\usepackage{dcolumn}% Align table columns on decimal point
\usepackage{bm}% bold math
\usepackage[normalem]{ulem} % For strikethrough
\usepackage{mathrsfs}
\usepackage{physics}
\usepackage{xcolor}
\usepackage{amsmath, amssymb, mathtools, amsthm}
\usepackage{comment} %create block comment
\usepackage{orcidlink}

\usepackage{bibunits}
\defaultbibliographystyle{apsrev4-2}

\newtheorem{theorem}{Theorem}
\graphicspath{{figs/}} %figures path
\begin{document}

% ========= MAIN TEXT =========
\begin{bibunit}
    \begin{abstract}
Understanding quantum tunneling in many-body systems is crucial for advancing quantum technologies and nanoscale device design. Despite extensive studies of quantum tunneling, the role of interactions in determining directional transport through asymmetric barriers in discrete quantum systems remains unclear. Here we show that noninteracting fermions exhibit symmetric tunneling probabilities regardless of barrier orientation, while inter-particle interactions break this symmetry and create pronounced asymmetric tunneling behavior. We explore the dependence of tunneling behavior on the initial spin configurations of two spin-1/2 fermions: spin-triplet states preserve tunneling symmetry, while spin-singlet states show strong asymmetry. We identify regimes where interactions mediate tunneling through under-barrier resonant trapping and enhance tunneling via many-body resonant tunneling -- a phenomenon arising solely from inter-particle interactions and being fundamentally different from traditional single-particle resonant tunneling. Our results may be applied to the design of nanoscale devices with tailored transport properties, such as diodes and memristors.
\end{abstract}

\title{Few-fermion resonant tunneling and underbarrier trapping in asymmetric potentials}

\author{Elvira Bilokon\orcidlink{0009-0007-8296-2906}}
\email{ebilokon@tulane.edu}
\affiliation{Department of Physics and Engineering Physics, Tulane University, New Orleans, Louisiana 70118, United States}
\affiliation{Akhiezer Institute for Theoretical Physics, NSC KIPT, Akademichna 1, 61108 Kharkiv, Ukraine}

\author{Valeriia Bilokon\orcidlink{0009-0001-1891-0171}}
\email{vbilokon@tulane.edu}
\affiliation{Department of Physics and Engineering Physics, Tulane University, New Orleans, Louisiana 70118, United States}
\affiliation{Akhiezer Institute for Theoretical Physics, NSC KIPT, Akademichna 1, 61108 Kharkiv, Ukraine}

\author{Dusty R. Lindberg\orcidlink{0001-5335-7941}}
\email{dlindberg@tulane.edu}
\affiliation{Department of Physics and Engineering Physics, Tulane University, New Orleans, Louisiana 70118, United States}

\author{Andrii Sotnikov\orcidlink{0000-0002-3632-4790}}
\email{a\_sotnikov@kipt.kharkov.ua}
\affiliation{Akhiezer Institute for Theoretical Physics, NSC KIPT, Akademichna 1, 61108 Kharkiv, Ukraine}
\affiliation{Karazin Kharkiv National University, Svobody Square 4, 61022 Kharkiv, Ukraine}

\author{Lev Kaplan\orcidlink{0000-0002-7256-3203}}
\email{lkaplan@tulane.edu}
\affiliation{Department of Physics and Engineering Physics, Tulane University, New Orleans, Louisiana 70118, United States}

\author{Denys I. Bondar \orcidlink{0000-0002-3626-4804}}
\email{dbondar@tulane.edu}
\affiliation{Department of Physics and Engineering Physics, Tulane University, New Orleans, Louisiana 70118, United States}

\maketitle

\section{Introduction} 
Tunneling is a pivotal phenomenon in quantum physics, underpinning an exceptionally broad range of applications. It plays a major role in nuclear fusion~\cite{Balantekin1998}, facilitates ionization under the influence of strong laser fields in atomic and molecular physics~\cite{Amini2019}, and explains transport effects in condensed matter physics~\cite{Liang2014}. In semiconductor technology, tunneling is exploited to create components such as tunnel diodes~\cite{Guzun2013}. It is also used in Josephson junctions~\cite{Albiez2005}, which are key ingredients in superconducting electronics, and in scanning tunneling microscopy~\cite{Bai2000} for atomic-scale surface imaging. Tunneling is crucial in black hole physics, particularly in the context of Hawking radiation~\cite{Arzano2005}, as well as in theories describing the early universe~\cite{Atkatz1994}.

In particular, tunneling plays a critical role in many-body physics. However, the study of the quantum many-body systems becomes exceedingly difficult in the presence of strong correlations. For better understanding, an effective approach is to study few-body systems, which capture the microscopic mechanisms underlying physical phenomena, such as tunneling dynamics, correlation effects, and quantum entanglement.

There are many theoretical studies on few-body systems, reflecting their importance in simplifying and understanding the complex behaviors of many-body quantum systems. Depending on system parameters, different ground-state configurations of one-dimensional clusters of fermionic atoms were found in Ref.~\cite{Gordillo2018}. Within the fermionization limit, Refs.~\cite{Lindgren2014} and \cite{Gordillo2017} explored the phenomena of ferromagnetism and the Mott insulator, respectively. The dynamics of one-dimensional few-body systems was studied in Refs.~\cite{Erdmann2018, Sowinski2013, Ishmukhamedov2019, Zollner2008_1, Zollner2008, Nandy2022, Koutentakis2020, Sowinski2021, Esslinger2005, Rontani2012, Brugger2025, Becker2023}. In particular, for spin-polarized fermionic systems, the overall dynamics depending on the mass ratio and the height of a double well barrier was discussed in Ref.~\cite{Erdmann2018}. The study~\cite{Sowinski2013} showed the role of degeneracy in a system with strong repulsive interactions, while others~\cite{Ishmukhamedov2019,Zollner2008_1, Zollner2008} focused on the transition from uncorrelated to correlated pair tunneling. 

Experimental realizations have demonstrated precise control over systems using ultracold atoms~\cite{Folling2007,Trotzky2008,Spielman2007,Zwierlein2004,Partridge2006, Shin2006}, enabling the observation of interaction-induced shifts~\cite{Folling2007,Serwane2011} and the realization of Mott-insulating states~\cite{Murmann2015}. In Ref.~\cite{Mukherjee2016}, the authors observed pair tunneling in the strongly interacting regime in an experiment simulating the one-dimensional two-particle Hubbard model. Other experiments demonstrated intricate behaviors and correlation effects in few-body systems~\cite{Zurn2012,Zurn2013}. In Ref.~\cite{Wenz2013}, the authors investigated the crossover from few- to many-body physics, providing insights into how interactions scale with particle number. Beyond atomic physics, the ability to control tunneling dynamics has far-reaching implications for next-generation technologies. From quantum computing to low-power electronics, precise manipulation of tunneling opens new frontiers in device engineering. The study of asymmetric tunneling could inspire the design of nanoscale devices where electron transport can be controlled by inter-particle interactions. For example, interaction-driven tunneling could play a pivotal role in next-generation transistor technologies, such as tunnel field-effect transistors~\cite{Zhang2017, Liu2025} and quantum field-effect transistors~\cite{Nadeem2021, Chuang2015}, where precisely controlled tunneling could improve device efficiency and enable new functionalities in low-power electronics. In addition, memristors could exploit asymmetric tunneling effects for precise control over resistance through charge transport mechanisms~\cite{Lin2023}. Similarly, interaction-driven tunneling asymmetry may enhance the sensitivity of nanoscale sensors that rely on charge transport properties.

In this paper, we theoretically study tunneling dynamics of a few-body fermionic system in the presence of an external asymmetric potential. While it is well established that noninteracting particles in infinite continuous one-dimensional space undergo symmetric tunneling regardless of barrier shape~\cite[Sec. 25]{Landau1981} \cite{shegelski_equal_2020}, we extend this understanding by proving that tunneling probability remains symmetric even in finite discrete systems. Significantly, we demonstrate that a conventional two-particle interaction induces the quantum phenomenon of asymmetric tunneling.  While advancing our understanding of the left-right tunneling symmetry breaking~\cite{Lindberg2023}, we unexpectedly open a treasure chest of interaction-induced quantum effects such as the nontrivial dependence of tunneling probability on the spin configuration of the initial state, with the triplet (singlet) state yielding symmetric (asymmetric) tunneling; underbarrier resonant trapping (a related effect in bosons was studied in Ref.~\cite{Kolovsky2012}); and many-body resonant tunneling. Note that the latter arises exclusively from inter-particle interactions and has nothing in common with the more widely known phenomenon of single-particle resonant tunneling~\cite{mohsen2013quantum}, which underpins the operation of resonant-tunneling diodes and even has classical analogues~\cite{dragoman_quantum-classical_2010}.
\section{Results} 

\subsection{Model and methods}\label{sec:model_methods} 

In classical physics, a particle cannot penetrate a potential barrier if its energy is lower than the barrier's potential energy. However, quantum mechanics introduces the concept of tunneling, allowing a particle to pass through the barrier with non-zero probability. Moreover, as shown by Landau~\cite[Sec. 25]{Landau1981} (see also~\cite{shegelski_equal_2020}), in a one-dimensional system, described by the time-independent Schr\"odinger equation, the probability of tunneling remains the same regardless of which side of the barrier the particle meets first. The theorem, however, is derived for continuous systems in infinite space and its applicability to discrete finite systems requires significant modifications. Moreover, the above also does not work for complex systems. In particular, many-body interactions lead to breaking of the tunneling symmetry~\cite{Amirkhanov1966, bondar_enhancement_2010, Lindberg2023}. 

In our theoretical approach, we examine tunneling dynamics in a one-dimensional discrete lattice system, focusing on 
the evolution generated by the paradigmatic Fermi--Hubbard Hamiltonian
\begin{eqnarray}\label{eq:FHM}
    \hat{\mathcal{H}} &=& -J \sum_{j=1, \sigma = \uparrow, \downarrow}^{L-1} \left( \hat{c}_{j, \sigma}^{\dagger} \hat{c}^{}_{j+1, \sigma} + \hat{c}_{j+1, \sigma}^{\dagger} \hat{c}^{}_{j, \sigma} \right) \\
    \nonumber
    &&+\sum_{j=1}^L (U\hat{n}_{j, \uparrow} \hat{n}_{j, \downarrow}+V^{\rm ex}_j\hat{n}_{j}) \,.
\end{eqnarray}
Here $\sigma=\uparrow, \downarrow$ denotes the spin, and the operator $\hat{c}_{j, \sigma}^{\dagger} (\hat{c}^{}_{j, \sigma})$ creates (annihilates) a fermion in the state $|\sigma_j \rangle$ at site~$j$. The number operator is defined as $\hat{n}_{j, \sigma}=\hat{c}_{j, \sigma}^{\dagger}\hat{c}^{}_{j, \sigma}$ and $\hat{n}_j=\hat{n}_{j, \uparrow}+\hat{n}_{j, \downarrow}$. The first term in Eq.~\eqref{eq:FHM} describes the kinetic energy with $J$ representing the hopping amplitude, while the second term accounts for the onsite interaction with strength $U$ and the site-dependent asymmetric external potential $V^{\rm ex}_{j}$. Namely, $V^{\rm ex}_{j}$ is defined as
\begin{align}\label{eq:ext_pot}
    V^{\rm ex~(a)}_{j} &=&
    \begin{cases}
        h,     & \text{if } j = L/2; \\
        h/2,   & \text{if } j = L/2+1; \\
        0,     & \text{otherwise};
    \end{cases}
    \\ \nonumber \mbox{ or } \\ \nonumber
    V^{\rm ex~(b)}_{j}&=&
    \begin{cases}
        h/2,     & \text{if } j = L/2; \\
        h,   & \text{if } j = L/2+1; \\
        0,     & \text{otherwise}.
    \end{cases}
\end{align}

It is worth mentioning that the external potential, hopping, and interaction terms are spin-independent. Expressing the components of operator $\mathbf{\hat{S}} = (\hat{S}_x, \hat{S}_y, \hat{S}_z)$ in terms of spin-1/2 Pauli matrices $\sigma^r$ as $\hat{S_r} = \frac{1}{2}\sum_{j=1}^L \hat{c}_{j\alpha}^\dagger \sigma^r_{\alpha\beta} \hat{c}^{}_{j\beta}$ for $\mathbf{r} = (x, y, z)$, one can ensure that the total spin operator $\hat{S}^2=\hat{S}_x^2 + \hat{S}_y^2 + \hat{S}_z^2$ commutes with $\mathcal{\hat{H}}$, i.e., $[\mathcal{\hat{H}},\hat{S}^2]=0$ (here and below we set the Planck constant $\hbar=1$). Hence, the system preserves the total spin, as it remains invariant under global SU(2) spin rotations.

\begin{figure}
%\begin{center}
    \includegraphics[width=\linewidth]{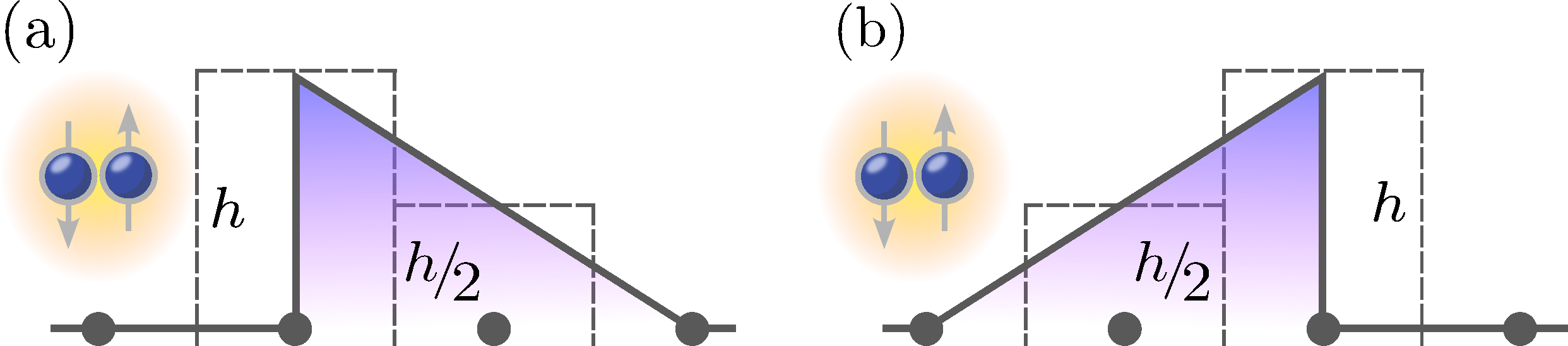} 
    \caption{\textbf{Initial configurations of the system with an asymmetric potential.} Schematic representation of two possible initial configurations of the interacting four-site system in the presence of an asymmetric external potential defined in Eq.~\eqref{eq:ext_pot}. While tunneling, particles first face (a) the barrier of height $h$; or (b) a gradual increase of the potential. Gray circles indicate the lattice sites. Dashed lines correspond to the actual shape of the barrier simulated in the numerical calculations, whereas solid lines are used to illustrate the triangular barrier shape.}
    \label{fig:in_config}
%\end{center}
\end{figure}

The smallest nontrivial system size is $L=4$. As shown in Fig.~\ref{fig:in_config}, two middle sites experience an asymmetric ``triangular'' (or a single reduced element of a ``sawtooth'') potential with maximum height~$h$. This configuration allows us to create two different initial states, since the Pauli exclusion principle allows for a maximum of two fermions on one side of the barrier. In the first case, see Fig.~\ref{fig:in_config}(a), two particles with opposite spins (a doublon) are positioned before the steep side of the barrier. Respectively, Fig.~\ref{fig:in_config}(b) demonstrates the second possible initial state with the two particles residing in front of the gradually increasing potential. The quantity under study is the expectation value of the number operator at the site after the barrier. 

As previously mentioned, ultracold atom experiments provide a highly controllable platform for realizing this system \cite{Folling2007, Trotzky2008, Spielman2007, Zwierlein2004, Partridge2006, Shin2006}, enabling precise tuning of the tunneling amplitude \cite{Esslinger2015, Spielman2012, Murmann2015} and interaction strength \cite{Chin2010, Ketterle2005, Walraven2010, Grimm2023}.

All our calculations are performed using the QuSpin package~\cite{Weinberg2017, Weinberg2019}. The latter is an open-source Python package designed for simulating the quantum dynamics of many-body systems. In particular, it provides the exact diagonalization (ED) method and employs a truncated Taylor series expansion to compute the action of the Hamiltonian's matrix exponential for time evolution. Note that QuSpin utilizes pseudo-spinors.

\subsection{Noninteracting fermions}\label{sec:nonint} 
\begin{figure}
%\begin{center}
    \includegraphics[width=\linewidth]{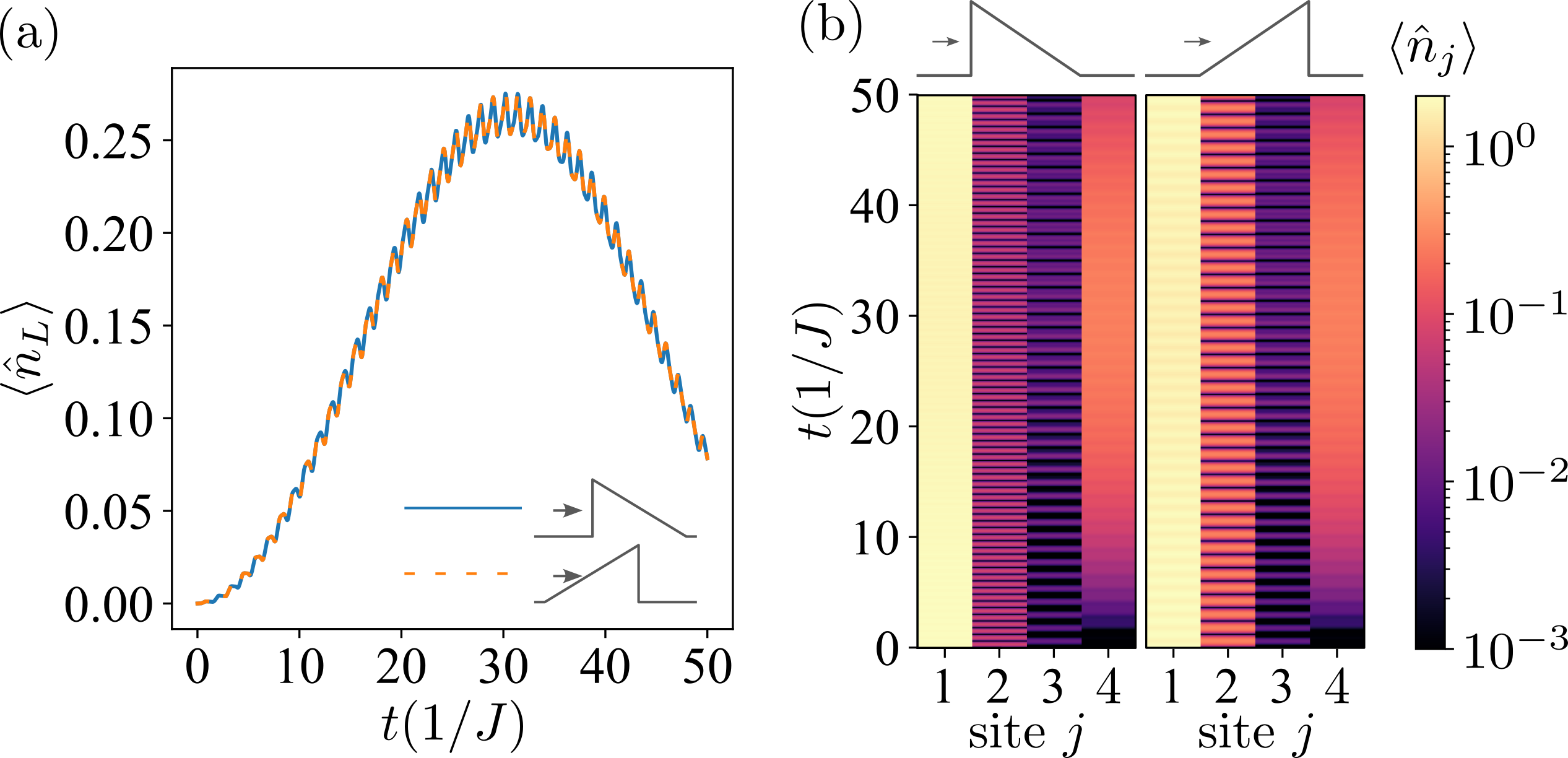}%{figs/U0_one_line.png}%
    \caption{\textbf{Time evolution of the number operator for the system in the initial noninteracting doublon state.} For the $L=4$ system {with a noninteracting ($U=0$) doublon as an initial state}, the expectation value of (a) $\langle \hat{n}_L\rangle$, and (b) $\langle \hat{n}_j\rangle$ (with values color-coded on a logarithmic scale) are shown as a function of time. The height of the barrier is $h=10 J$.}
    \label{fig:zero_U}
%\end{center}
\end{figure}

We begin by considering noninteracting fermions. In the case $U=0$, the many-body problem under study reduces to a single-particle problem. Hence, one can expect the tunneling rate to remain unchanged regardless of the barrier orientation. 
 To show this, we calculate the expectation value of the number operator on the last site $\hat{n}_L$, i.e., on the fourth site for the $L=4$ system. The blue curve and the orange dashed line in Fig.~\ref{fig:zero_U} correspond to the case of the system initially prepared in the state illustrated in Fig.~\ref{fig:in_config}(a) and Fig.~\ref{fig:in_config}(b), respectively. Although the tunneling particles face barrier sides of different height in these configurations, the two curves overlap perfectly. The primary contribution to the observable~$\langle \hat{n}_L \rangle$  comes from the tunneling of the single particle after doublon dissociation, and not from the tunneling of a doublon as a whole (the corresponding two-body contributions are negligibly small). Note that timescales considered in our simulations are within the experimentally accessible range in cold-atom lattice systems \cite{Greiner2016, Bakr2019, Zwierlein2019}. 
 
The observation of the symmetric tunneling turns out to be an illustration of the theorem, rigorously proven in Supplementary Note 2, whose simplified formulation reads as
\begin{quote}
    For a one-dimensional discrete system of noninteracting particles in the presence of an asymmetric external potential, the probability of finding a particle on the other side of the barrier is independent of the side on which the initial wave function is localized.
\end{quote} 
Unlike the case considered by Landau~\cite[Sec. 25]{Landau1981} (see also \cite{shegelski_equal_2020}),  who analyzed one-dimensional time-independent Schrödinger equation for a single particle -- which is a second-order ordinary differential equation -- our proof involves a finite discrete system, therefore relying on entirely different mathematical techniques. The key idea behind our derivation is that for every possible Feynman quantum path that a particle can take while tunneling in one direction, there exists a corresponding ``twin'' path for tunneling in the opposite direction that has exactly the same amplitude.
Since our theorem considers a single-particle system, its validity is independent of the particle statistics.

Despite the symmetric behavior of $\langle \hat{n}_L\rangle$, the full dynamics of the system is nontrivial, as seen when we calculate the expectation value of the number operator as a function of time for each site individually [see Fig.~\ref{fig:zero_U}(b), where the values of $\langle \hat{n}_j\rangle$ are color-coded]. We see that although the tunneling probability is symmetric, the dynamics of the entire system is asymmetric even in the absence of interaction.

\subsection{Symmetry considerations}\label{sec:sym_cons} 
\begin{figure}
\begin{center}
    \includegraphics[width=\linewidth]{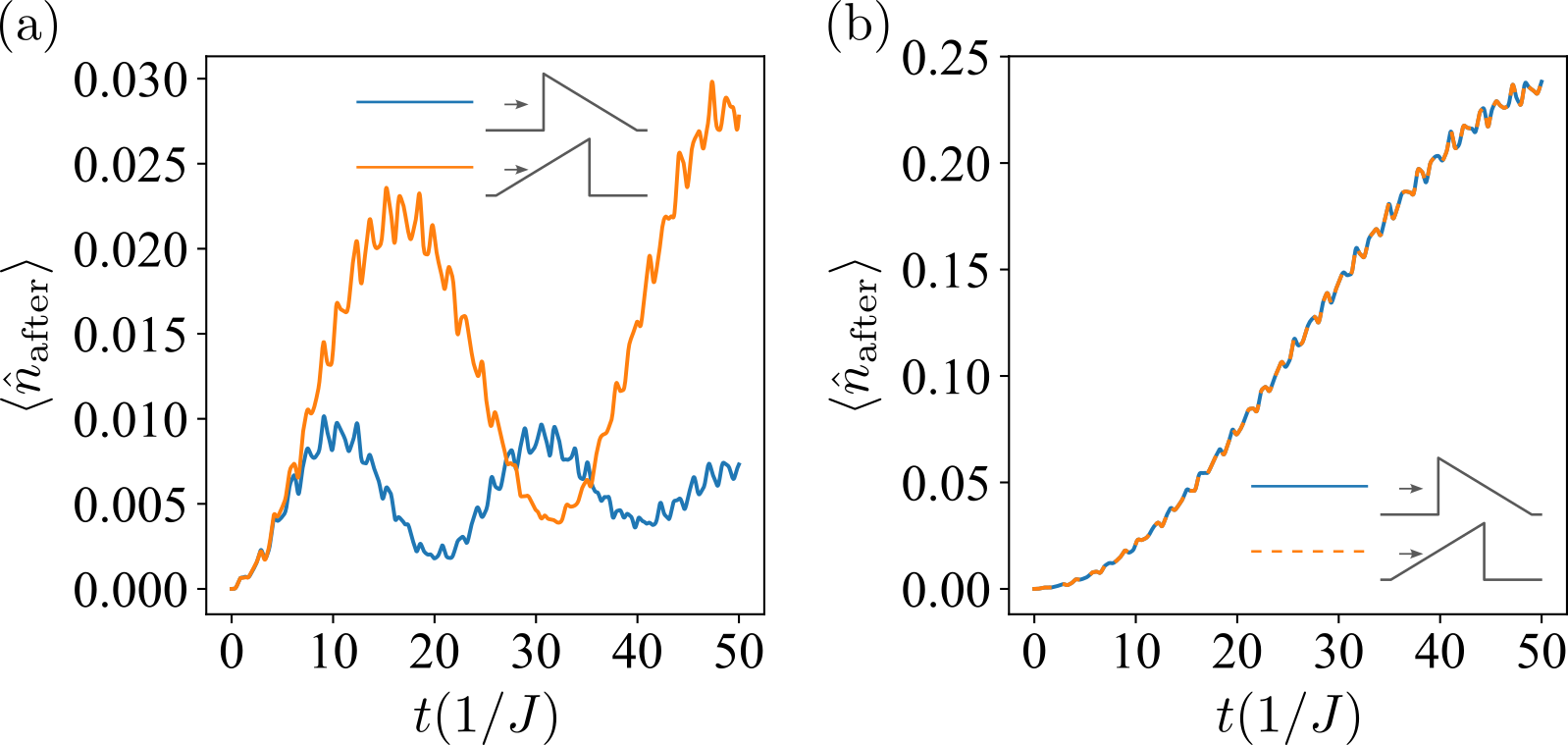}
    \caption{\textbf{Time evolution of $\langle \hat{n}_{\rm after}\rangle$ for different initial spin states.} For the $L=6$ system, the expectation 
 value  $\langle \hat{n}_{\rm after} \rangle$ as a function of time is shown for (a) an initial singlet state, and (b) an initial triplet state. In both cases, the amplitude of the on-site interaction is $U=0.5 J$, and the height of the barrier is $h=10 J$.}
    \label{fig:triplet_singlet}
\end{center}
\end{figure}

Let us discuss the dynamics of the interacting system. As stated in Introduction, the interaction between particles induces the asymmetric tunneling effect (see also Supplementary Note 1). To demonstrate this, let us consider a $L=6$ system with two fermions prepared in the spin singlet initial state, $|\Psi_{\rm s}(0)\rangle=\frac{1}{\sqrt{2}}(|\uparrow_1\downarrow_2\rangle-|\downarrow_1\uparrow_2\rangle)$. One should bear in mind that regardless of the system size, we set the potential to act on the two central sites of the lattice, i.e., on the third and the fourth sites for $L=6$. Figure~\ref{fig:triplet_singlet}(a) depicts the evolution of $\langle \hat{n}_{\rm after} \rangle$, where $\hat{n}_{\rm after}$  represents the sum of the number operators acting on the sites after the barrier, i.e., $\langle \hat{n}_{\rm after} \rangle=\sum_{j=L/2+2}^L \langle\hat{n}_j \rangle$.
The clear difference between two curves indicates that the tunneling probability for the spin singlet configuration is asymmetric.
Alternatively, we can consider the system initially localized in the spin triplet state, 
$|\Psi_{\rm t}(0)\rangle=\frac{1}{\sqrt{2}}(|\uparrow_1\downarrow_2\rangle+|\downarrow_1\uparrow_2\rangle)$. Fig.~\ref{fig:triplet_singlet}(b) demonstrates the evolution of $\langle\hat{n}_{\rm after}\rangle$. Since two curves fully align throughout the entire time range, the initial triplet state reveals symmetric tunneling probability, as in the noninteracting system. The system size for this part of the discussion is chosen based on the smallest possible configuration that allows for the positioning of particles in the triplet state. Note that when employing a numerical package for simulating quantum dynamics (see Sec.~\ref{sec:model_methods}), one should carefully check whether the numerical method employs a spinor or pseudo-spinor basis in order to correctly attribute the singlet and triplet character to the states. For the reader's convenience, we define the spin-singlet (spin-triplet) state within the spinor basis.

To understand the dynamics described above, one can consider the action of the time evolution operator on the initial state,
\begin{equation}\label{eq:time_evol_expansion}
    e^{-i\hat{\mathcal{H}}t}|\Psi(0)\rangle=\left(1-i\hat{\mathcal{H}}t+\frac{1}{2}(-i\hat{\mathcal{H}}t)^2+...\right)|\Psi(0)\rangle \,.
\end{equation}
Writing down explicitly the first two terms of the expansion, the action on the singlet state reads
\begin{eqnarray}
    &&
    e^{-i\hat{\mathcal{H}}t}|\Psi_{\rm s}(0)\rangle = |\Psi_{\rm s}(0)\rangle \\ \nonumber 
    &-& it\left[\frac{-J}{\sqrt{2}}(2|\uparrow_1\downarrow_1\rangle+2|\uparrow_2\downarrow_2\rangle+|\uparrow_1\downarrow_3\rangle-|\downarrow_1\uparrow_3\rangle) \right] + \cdots \,.
\end{eqnarray}
We see that the linear term~($-i\hat{\mathcal{H}}t$) in Eq.~\eqref{eq:time_evol_expansion} creates states with two particles occupying the same site. Hence, the action of $\frac{1}{2}(-i\hat{\mathcal{H}}t)^2$ produces terms involving the on-site interaction with strength $U$. At the same time, for a triplet initial state, the terms describing doubly occupied sites cancel:
\begin{eqnarray}
    &&e^{-i\hat{\mathcal{H}}t}|\Psi_{\rm t}(0)\rangle = 
    |\Psi_{\rm t}(0)\rangle 
    \\ \nonumber 
    &-& it\left[\frac{-J}{\sqrt{2}}(|\uparrow_1\downarrow_3\rangle+|\downarrow_1\uparrow_3\rangle) \right] + \cdots \,.
\end{eqnarray}
Moreover, such cancellation of the interaction-dependent terms also takes place in the higher-order terms in the expansion \eqref{eq:time_evol_expansion}. Hence, by prohibiting formation of doublons (with $\hat{S}^2 \ket{\uparrow_1\downarrow_1}=0$), a triplet state preserves the symmetry of tunneling probability. This also agrees with the arguments based on the preservation of the total spin in the system, as pointed out in the Sec.~\ref{sec:model_methods}.

\section{Nontrivial dynamics in asymmetric potentials}
\begin{figure}
\begin{center}
    \includegraphics[width=\linewidth]{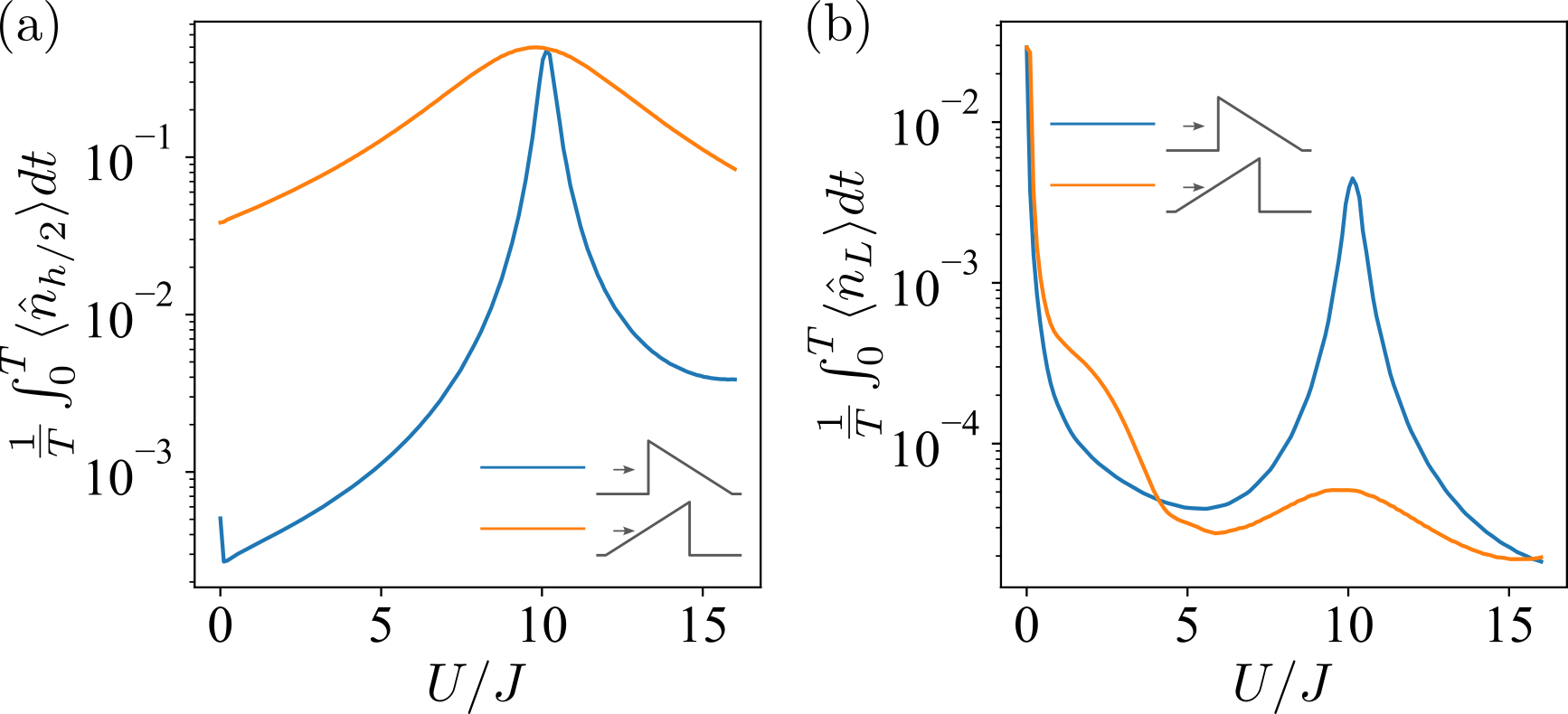}
    \caption{\textbf{Time-averaged expectation value of the number operator as a function of interaction strength.} For the $L=4$ system, the time-averaged expectation value of (a) $\langle\hat{n}_{h/2}\rangle$, and (b) $\langle\hat{n}_{L}\rangle$  as a function of the interaction amplitude $U$ (in units of $J$). The initial state is a doublon positioned at the first site. The barrier height is $h=20 J$ and $T=100/J$. For $U=10 J$, the interaction strength coincides with the external potential strength at the site $j^*$.}
    \label{fig:dens_avg}
\end{center}
\end{figure}

{Let $j^\ast$ denote the lattice site $j$ with external potential $V_j^{\rm ex}=h/2$ and $\hat{n}_{h/2}$ denote the number operator acting on $j^\ast$.}
For $h=20 J$, the time-averaged (over the interval {$T=100/J$}
) expectation values of $\hat{n}_{h/2}$ and $\hat{n}_L$ are depicted in Fig.~\ref{fig:dens_avg}. The initial state is represented by a doublon positioned at the first site. As one can see from Fig.~\ref{fig:dens_avg}(a), $\langle\hat{n}_{h/2}\rangle$ is larger in the case of tunneling from the angled side of the barrier. This can be explained by the fact that particles placed on this side require fewer evolution steps
to reach the lattice site $j^\ast$. The results shown in Fig.~\ref{fig:dens_avg}(b) agree with the discussion of the noninteracting case, since the two $\hat{n}_L$ curves meet at $U=0$. However, as $U$ increases, the time-averaged values of $\hat{n}$ depend on the initial state. At $U\approx10J$, both $\langle\hat{n}_L\rangle$ and $\langle\hat{n}_{h/2}\rangle$ show noticeable growth. The behavior in this parameter regime is discussed below and is largely unaffected by increasing the final time $T$. 

\subsection{Underbarrier resonant trapping}\label{sec:res_trap}
\begin{figure}
\begin{center}    \includegraphics[width=0.75\linewidth]{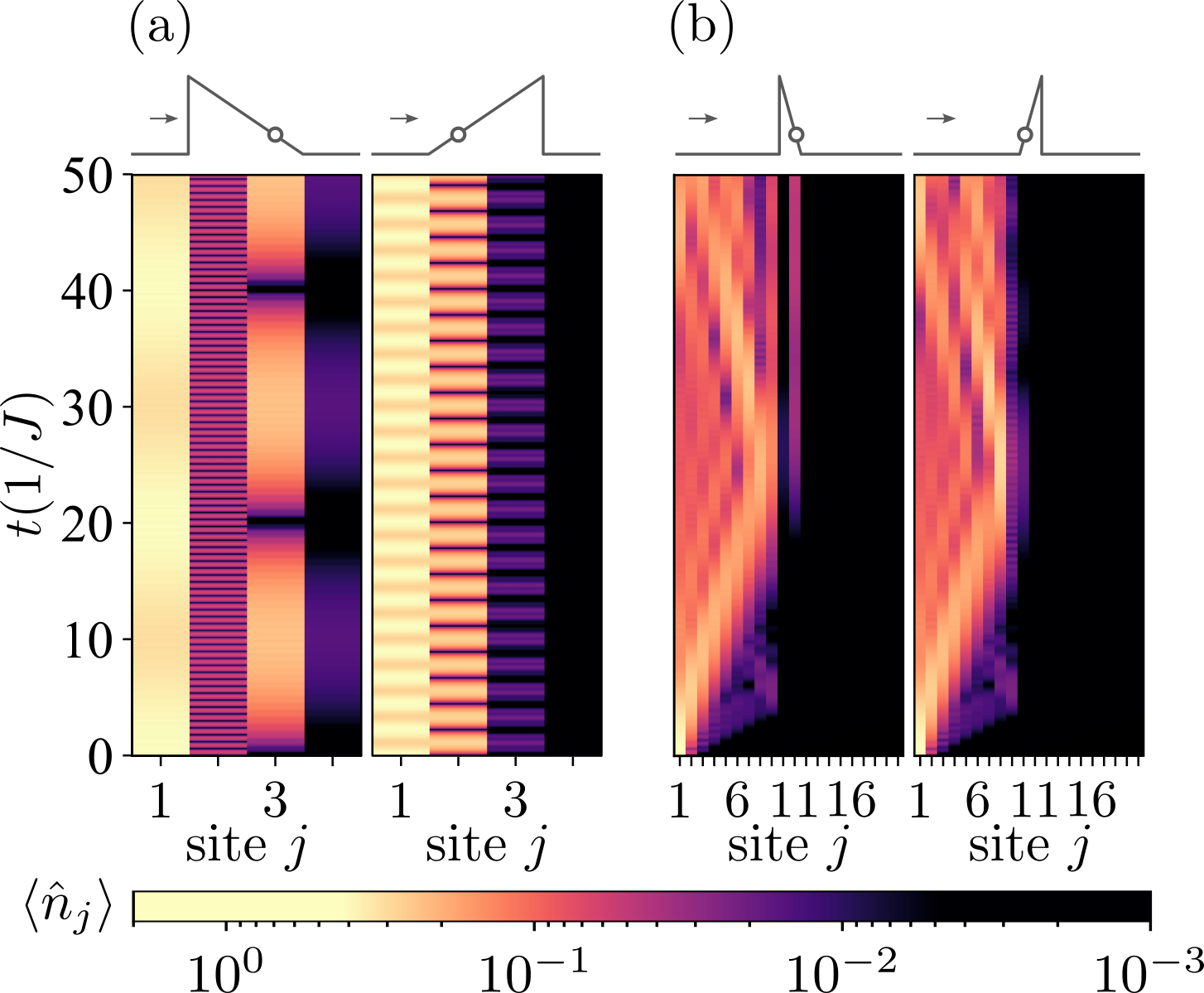}
\caption{\textbf{Phenomenon of the underbarrier resonant trapping.} For different system sizes, $L=4$ (a) and $L=20$ (b), the evolution of $\langle \hat{n}_j \rangle$ is shown color-coded on a logarithmic scale. The initial state is a doublon positioned at the first site. The interaction strength is $U=10 J$. The small schematic illustrations above each panel represent the potential barrier with circles indicating the site $j^\ast$ where the potential barrier satisfies the resonant condition $h/2=U$.}
    \label{fig:res_trap}
\end{center}
\end{figure}

In the following, let us fix the interaction strength $U$ to be half of the barrier height $h$, i.e., $U=h/2$. Figure~\ref{fig:res_trap} illustrates the evolution of the particle density $\langle \hat{n}_j \rangle$ for lattice systems with $L=4$ (a) and $L=20$ (b).  As shown, 
there is a significant increase in $\langle \hat{n}_{h/2}\rangle$, i.e., the occupation of the site~$j^\ast$.
We refer to this phenomenon as underbarrier resonant trapping. Note that the populations on the other side of the barrier are extremely small, indicating a suppression of tunneling. This trapping results from the conservation of total energy. The energy of the initial doublon state is $U$. To conserve this energy, one particle can remain before the barrier with negligible kinetic energy, while the other is trapped at the site~$j^\ast$ with the potential energy $h/2=U$. It is worth mentioning that the phenomenon of the underbarrier resonant trapping is robust with respect to the increase in the system size. However, larger systems introduce more degrees of freedom and require more evolution steps to reach the site $j^\ast$. As the result, the increased complexity requires a longer time to observe the underbarrier resonant trapping (see Supplementary Note 3).

\subsection{Highly-asymmetric resonant tunneling}\label{sec:res_tun} 
\begin{figure}
\begin{center}
    \includegraphics[width=\linewidth]{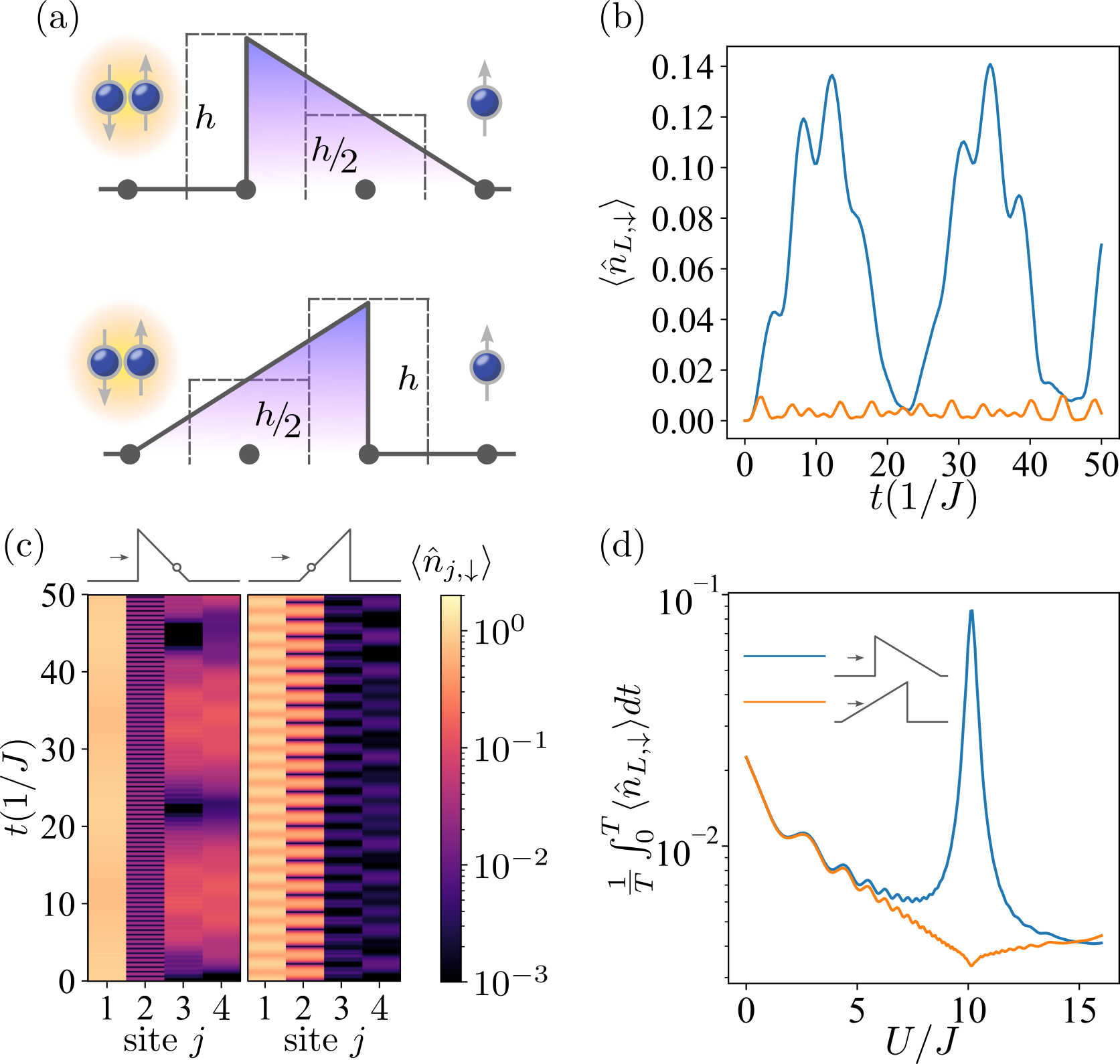}
    \caption{\textbf{Phenomenon of the highly-asymmetric resonant tunneling.} For the $L=4$ system with $h=20 J$, (a) a doublon initially placed on the site before the barrier and a spin-$\uparrow$ fermion placed after it. We show the expectation values of (b) $\langle \hat{n}_{L, \downarrow}\rangle$, and (c) $\langle \hat{n}_{j,\downarrow}\rangle$ (with values color-coded on a logarithmic scale) as a function of time. (d) The time-averaged expectation value of $\langle \hat{n}_{L,\downarrow}\rangle$ (with $T=100/J$) as a function of the interaction amplitude $U$. The circle in the schematic barrier in (c) illustration indicates the site $j^\ast$ where the potential is $h/2=U$.}
    \label{fig:res_tun}
\end{center}
\end{figure}

Another effect taking place when $U=h/2$ is the phenomenon of many-body resonant tunneling. As an example, consider the $L=4$ system with three particles: a doublon and a spin-up particle on opposite edges of the one-dimensional lattice, separated by the two-site asymmetric barrier in the center, see Fig.~\ref{fig:res_tun}(a). 
We expect that the presence of one additional particle after the barrier can cause an enhancement of tunneling. Namely, during the evolution for $U=h/2$, energy conservation can be achieved not only by trapping the fermion at the site~$j^\ast$, 
but also by a doublon formation after the barrier. Figure~\ref{fig:res_tun}(b) shows the time evolution of $\langle \hat{n}_{L,\downarrow}\rangle$. As one can see, in the case where the doublon faces the angled side of the barrier, the respective density values beyond the barrier are negligibly small. In fact, it is clear from Fig.~\ref{fig:res_tun}(c) that the spin-$\downarrow$ particle barely penetrates the barrier. The behavior changes drastically when the doublon is facing the steep side of the potential. As depicted in Fig.~\ref{fig:res_tun}(b), the values of $\langle \hat{n}_{L,\downarrow}\rangle$ are an order magnitude higher when the spin-$\downarrow$ fermion tunnels from the steep side. This illustrates that the presence of the additional spin-$\uparrow$ particle after the barrier does enhance the tunneling [see the order of magnitude of the average tunneling probability in Fig.~\ref{fig:res_tun}(d) and the corresponding values in Fig.~\ref{fig:dens_avg}(b)].

The stark contrast in tunneling probabilities shown in Fig.~\ref{fig:res_tun}(b) is explained as follows: Due to conservation of energy, the tunneling fermion can exit the barrier only as part of a doublon. It is very easy for such a doublon to form when the tunneling particle faces the sharp edge of the potential because underbarrier trapping brings the tunneling fermion into close proximity to the fermion initially located after the barrier. However, when initially facing the smooth ramp, underbarrier trapping leaves a gap between the tunneling fermion and its potential partner after the barrier. Hence, an extra tunneling step is required for them to approach each other and form the doublon.

In conclusion, we reiterate that the phenomenon of many-body quantum tunneling has no single-particle analogue.
In the single-particle case, the resonant tunneling cannot occur for a single barrier, requiring two or more barriers~\cite{mohsen2013quantum}. 
In contrast, our many-body resonant tunneling occurs in a single barrier and arises from the interplay between inter-particle interactions and the external potential.

\section{Conclusion} 
We have investigated tunneling dynamics in a discrete few-fermion system under the influence of an asymmetric external potential. Our analysis shows that, for noninteracting particles, the tunneling probability remains symmetric regardless of the barrier's orientation. However, the symmetry breaks down in the presence of inter-particle interactions, and the system evolution depends on the initial configuration.  

For slightly larger systems that provide more ways to distribute fermions in the initial state, we showed that the tunneling behavior reveals a dependence on the initial spin configurations. The spin-triplet state formed by two spin-1/2 fermions occupying nearest-neighbor sites preserves the symmetry of tunneling probability (aligning with the noninteracting behavior), whereas the spin-singlet state exhibits notable interaction-induced asymmetry in the tunneling probability.

Also, we explored the system for a specific ratio between the interaction strength $U$ and the potential strength $V^{\rm ex}_j$, i.e., $V^{\rm ex}_j=U=h/2$. We identified conditions under which the interaction causes underbarrier resonant trapping and conditions under which interaction enhances tunneling (the highly-asymmetric resonant tunneling regime). These phenomena underscore the complex interplay between interaction strength, barrier height, and initial state configuration in determining tunneling dynamics. Our study highlights the richness and tunable behavior of few-fermion systems in the presence of an asymmetric barrier and points toward possible applications in quantum control where interaction-driven and symmetry-breaking effects are essential.

\section*{Code availability}
The code is available on \href{https://github.com/ebilokon/1band_fermi_hubbard/tree/main/AsymmetricTunneling}{GitHub}. 

\acknowledgments
We sincerely thank the referees for their valuable comments and suggestions, which have significantly improved the clarity and quality of our manuscript. We are grateful to Prof.~Immanuel Bloch and Prof.~Artem Volosniev for inspiring discussions. We also thank Stanislava Litvinova for thorough verification of numerical results. This work was supported by the National Science Foundation (NSF) IMPRESS-U Grant No.~2403609. D.R.L. is supported by NASA EPSCoR and/or the Board of Regents Support Fund. A.S. acknowledges support by the National Research Foundation of Ukraine, project No.~0124U004372. D.I.B. is also supported by Army Research Office (ARO) (grant W911NF-23-1-0288; program manager Dr.~James Joseph). The views and conclusions contained in this document are those of the authors and should not be interpreted as representing the official policies, either expressed or implied, of ARO, NSF, or the U.S. Government. The U.S. Government is authorized to reproduce and distribute reprints for Government purposes notwithstanding any copyright notation herein.

    \putbib[fermions]
\end{bibunit}

% ========= SUPPLEMENT =========
\clearpage
\newcommand{\suppnote}[2]{%
  \vspace{1em}%
  \begin{center}
  \noindent\textnormal{\textbf{Supplementary Note #1: \MakeUppercase{#2}}\\[1ex]}%
  \end{center}
}

\renewcommand{\figurename}{Supplementary Figure}

\setcounter{equation}{0}
\renewcommand{\theequation}{\arabic{equation}}
\setcounter{figure}{0}
\renewcommand{\thefigure}{\arabic{figure}}

\begin{bibunit}
    \title{Supplementary Information for ``Few-fermion resonant tunneling and underbarrier trapping in asymmetric potentials''}

\maketitle

\suppnote{1}{Interacting fermions: {the Falicov--Kimball limit}}
As found, noninteracting particles undergo symmetric tunneling regardless of the barrier orientation. However, the system evolution changes significantly as soon as the interaction $U$ between fermions is included. To illustrate this, we consider the case of an immobile particle, which corresponds to the Falicov-Kimball limit. The Falicov-Kimball model~\cite{Falicov1969, Freericks2003} arises in condensed matter physics to describe correlated electron systems, particularly metal-insulator transitions in compounds with both itinerant and localized electrons. Similar to our model, the itinerant particles do not interact directly with each other but experience an interaction with the localized particles that are bound to specific locations in the lattice. This interaction can effectively act as a site-dependent potential barrier that influences the movement of itinerant electrons. With only one spin component allowed to propagate, the Falicov-Kimball model becomes simpler than the Hubbard model and is exactly solvable in the limit of infinite spatial dimensions~\cite{Lemanski2017, Kapcia2019}. 

{The Fermi--Hubbard Hamiltonian that describes our model reads
\begin{eqnarray}\label{eq:FHM}
    \hat{\mathcal{H}} &=& -J \sum_{j=1, \sigma = \uparrow, \downarrow}^{L-1} \left( \hat{c}_{j, \sigma}^{\dagger} \hat{c}^{}_{j+1, \sigma} + \hat{c}_{j+1, \sigma}^{\dagger} \hat{c}^{}_{j, \sigma} \right) \\
    \nonumber
    &&+\sum_{j=1}^L (U\hat{n}_{j, \uparrow} \hat{n}_{j, \downarrow}+V^{\rm ex}_j\hat{n}_{j}) 
\end{eqnarray}
with site-dependent asymmetric external potential $V^{\rm ex}_{j}$ defined as
\begin{align}\label{eq:ext_pot}
    &V^{\rm ex~(a)}_{j} =
    \begin{cases}
        h,     & \text{if } j = L/2; \\ \nonumber
        h/2,   & \text{if } j = L/2+1; \\
        0,     & \text{otherwise}; 
    \end{cases} \\   
    \mbox{ or } \\ \nonumber  
    &V^{\rm ex~(b)}_{j} =
    \begin{cases}
        h/2,     & \text{if } j = L/2; \\
        h,   & \text{if } j = L/2+1; \\
        0,     & \text{otherwise}.
    \end{cases}
\end{align}
}

In our case, a fermion with $\sigma=\downarrow$ acts as a localized particle. Since it remains confined to the first site, we only need to track the position of the itinerant fermion with $\sigma = \uparrow$.
As a result, the Hamiltonian from Supplementary Figure~\eqref{eq:FHM} becomes
\begin{eqnarray}\label{eq:FK_limit}
    \hat{\mathcal{H}} &=& -J \sum_{j=1}^{L-1} \left( \hat{c}_{j, \uparrow}^{\dagger} \hat{c}_{j+1, \uparrow} + \hat{c}_{j+1, \uparrow}^{\dagger} \hat{c}_{j, \uparrow} \right) \\
    \nonumber
    &+&\sum_{j=1}^L (U\hat{n}_{j, \uparrow} \delta_{j,1}+V^{\rm ex}_j\hat{n}_{j, \uparrow}) \,.
\end{eqnarray}
Explicitly, for the case when the particles face the vertical side of the barrier [see Supplementary Figure~\ref{fig:frozen}(a)], {$V^{\rm ex}_{L/2}=V^{\rm ex}_2=h$ and $V^{\rm ex}_{L/2+1}=V^{\rm ex}_3=h/2$. This leads to the following Hamiltonian}
\begin{eqnarray}\label{eq:Hfroz_vertical}
    \hat{\mathcal{H}}_a &=& -J \sum_{j=1}^{L-1} \left( \hat{c}_{j, \uparrow}^{\dagger} \hat{c}_{j+1, \uparrow} + \hat{c}_{j+1, \uparrow}^{\dagger} \hat{c}_{j, \uparrow} \right) \\
    \nonumber
    &+&U\hat{n}_{1,\uparrow}+h\hat{n}_{2,\uparrow}+\frac{h}{2}\hat{n}_{3,\uparrow}\,.
\end{eqnarray}
In the second case [see Supplementary Figure~\ref{fig:frozen}(b)], when particles are positioned before the angled side of the barrier, {$V^{\rm ex}_{L/2}=V^{\rm ex}_2=h/2$ and $V^{\rm ex}_{L/2+1}=V^{\rm ex}_3=h$. Then Supplementary Equation~\eqref{eq:FK_limit} transforms into} 
\begin{eqnarray}\label{eq:Hfroz_angled}
    \hat{\mathcal{H}}_b &=& -J \sum_{j=1}^{L-1} \left( \hat{c}_{j, \uparrow}^{\dagger} \hat{c}_{j+1, \uparrow} + \hat{c}_{j+1, \uparrow}^{\dagger} \hat{c}_{j, \uparrow} \right) \\
    \nonumber
    &+&U\hat{n}_{1,\uparrow}+\frac{h}{2}\hat{n}_{2,\uparrow}+h\hat{n}_{3,\uparrow}\,.
\end{eqnarray}
{It is clear from Supplementary Equations~\eqref{eq:Hfroz_vertical} and \eqref{eq:Hfroz_angled} that the inter-particle interaction $U$ effectively changes the external potential by altering the local energy of the site when a spin-up particle is present. As one can see from Supplementary Figure~\ref{fig:frozen}, the modified effective barriers, depicted by dashed lines, are not mirror images of each other.} Thus, the asymmetry in $\langle\hat{n}_L\rangle$ is caused by penetration of the spin-$\uparrow$ particle through two different barriers.

\begin{figure}
\begin{center}
    \includegraphics[width=0.9\linewidth]{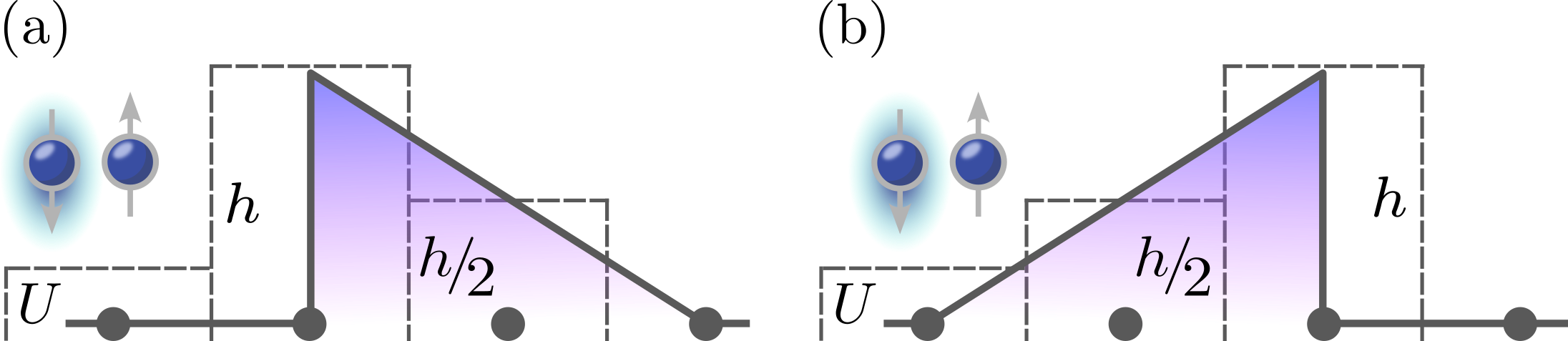}
    \caption{{\textbf{Initial configurations in the Falicov-Kimball limit.} Within Falicov-Kimball limit, spin-$\downarrow$ and spin-$\uparrow$ fermions represent the localized and itinerant particles, respectively. The interaction $U$ induces an effective modification of the external potential $V^{\rm ex}_{j}$. The modified effective barrier is depicted by dashed lines.}}
    \label{fig:frozen}
\end{center}
\end{figure}

\suppnote{2}{Proof of the symmetry of tunneling probability in the absence of interaction}

\begin{theorem}\label{RigorousTheorem}
Let us assume 
\begin{enumerate}
    \item The Fermi-Hubbard Hamiltonian $\hat{\mathcal{H}}$ from Supplementary Equation~\eqref{eq:FHM} with no interaction $U=0$ with the external potential $V^{\rm ex}_j$ schematically depicted in Supplementary Figure~\ref{fig:proof}; $V^{\rm ex}_j = 0$ for $j$ in regions $A$ and $C$, whereas  $V^{\rm ex}_j$ is arbitrary in region $B$. 
    \item The probability of a particle tunneling into region~$C$ is
    \begin{align}
        & \langle \hat{n}_C(t) \rangle = \bra{\Psi(t)}  \hat{n}_C \ket{\Psi(t)}, \notag\\
        & \hat{n}_C=\sum_{c=L_A+L_B+1}^L \hat{n}_c,
    \end{align}
    where the wavefunction $\ket{\Psi(t)} = e^{-it \hat{\mathcal{H}}} \ket{\Psi(0)}$ describes evolution of the initial state $\ket{\Psi(0)}$ of a single particle positioned in region~$A$, i.e.,
    \begin{align}
        & \ket{\Psi(0)}=\sum_{a=1}^{L_A} \kappa_a \hat{c}_a^\dagger|0\rangle, 
        \qquad \sum_{a=1}^{L_A} |\kappa_a|^2 =1.
    \end{align}
    \item The probability of a particle tunneling into region~$A$ is
    \begin{align}
        & \langle \hat{n}_A(t) \rangle = \bra{\widetilde{\Psi}(t)}  \hat{n}_A \ket{\widetilde{\Psi}(t)}, \notag\\
        & \hat{n}_A=\sum_{a=1}^{L_A} \hat{n}_a,
    \end{align}
     where the wavefunction $\ket{\widetilde{\Psi}(t)} = e^{-it \hat{\mathcal{H}}} \ket{\widetilde{\Psi}(0)}$ describes evolution of the initial state $\ket{\widetilde{\Psi}(0)}$ which is the mirror reflection of $\ket{\Psi(0)}$, i.e.,
     \begin{align}
             \ket{\widetilde{\Psi}(0)}=\sum_{a=1}^{L_A} \kappa_{a} \hat{c}_{L-a+1}^\dagger|0\rangle.
     \end{align}
     While $\ket{\Psi(0)}$ is localized in region $A$, the initial state $\ket{\widetilde{\Psi}(0)}$ is localized in region $C$.
\end{enumerate} 
Under these assumptions,
\begin{align}
    \langle \hat{n}_C(t) \rangle  = \langle \hat{n}_A(t) \rangle \mbox{ for all times $t$}.
\end{align}
In other words, the probability of tunneling from left to right equals the probability of tunneling from right to left for a single noninteracting particle initially localized before an arbitrary (including asymmetric) barrier.   
\end{theorem}

\begin{figure}
\begin{center}
    \includegraphics[width=0.65\linewidth]{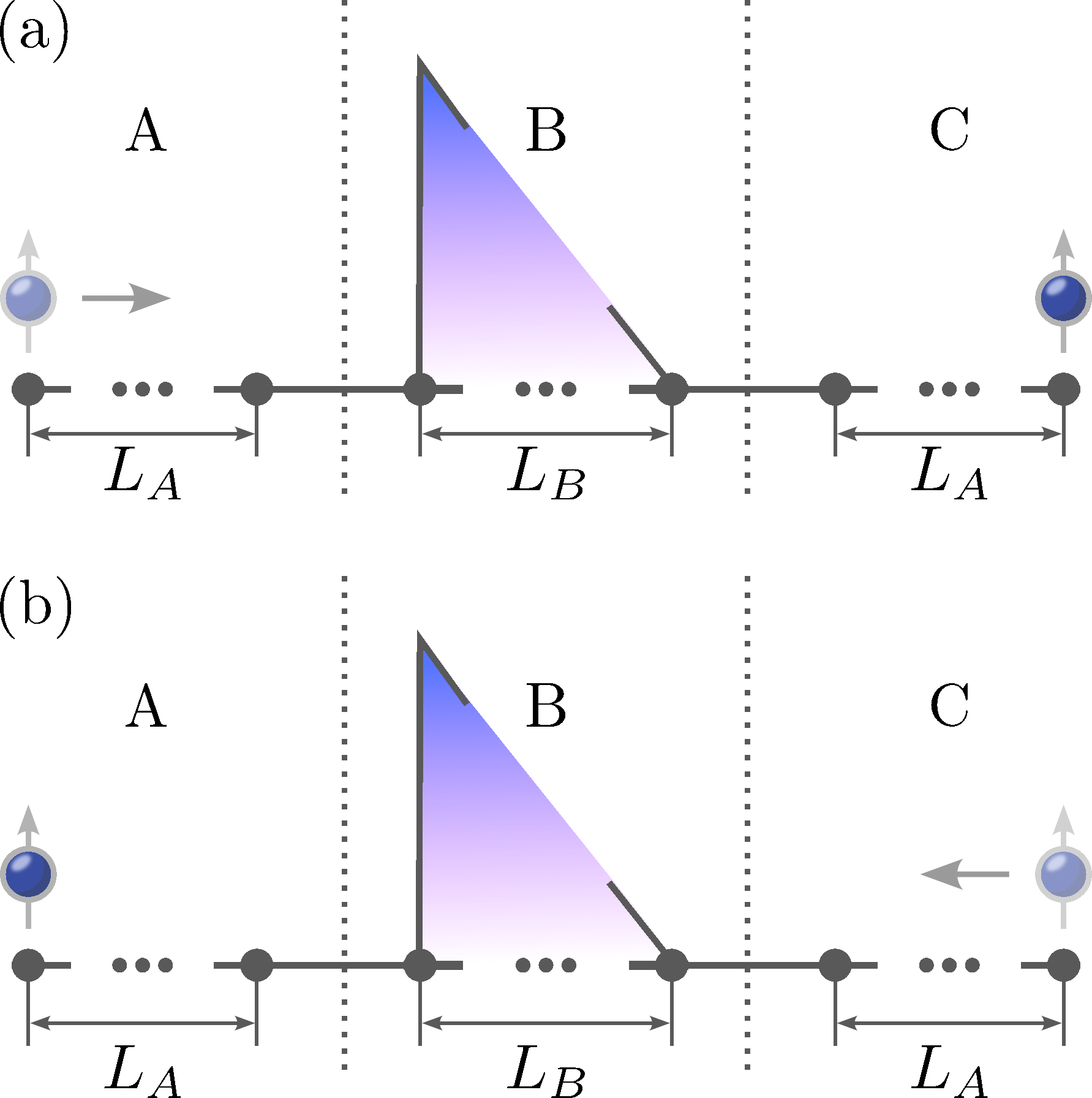}
    \caption{\textbf{System partition and the initial position of the particle.} For an arbitrary system size, the chain can be divided into three parts: $A$ -- the region before the barrier, $B$ -- the region with an asymmetric barrier, and $C$ -- the region after the barrier. The number of sites before and after the barrier is $L_A$, whereas $L_B$ is the number of sites under the action of the external potential. The particle is initially positioned (a) from the vertical side, and (b) from the angled side of the barrier. The gray arrow indicates the tunneling direction.}
    \label{fig:proof}
\end{center}
\end{figure}

\begin{proof}
For a single particle, the Hubbard Hamiltonian $\hat{\mathcal{H}}$  acts simply as an $L \times L$ matrix $H$, with 
\begin{equation}
    H_{i,j}=V^{\rm ex}_i \delta_{i,j} -J \delta_{i,j+1}-J \delta_{i,j-1} \,,
\end{equation}
and as indicated above $V^{\rm ex}_i$ vanishes in regions $A$ and $C$. For the following argument, the essential properties of $H$ are that {(i) $H$ is real,} (ii) the restriction of $H$ to region $C$ is identical, after reflection, to the restriction of $H$ to region $A$, and (iii) only nearest-neighbor hopping is present, so that there is a single coupling linking regions $A$ and $B$ and a single (identical) coupling linking regions $C$ and $B$. 

We start with an amplitude $\langle c |e^{-iH t}|a\rangle$, which we want to relate to $\langle L+1-c |e^{-iH t}|L+1-a\rangle$, where $|a\rangle$ and $|c\rangle$ are in region $A$ and $C$, respectively (and $|L+1-a\rangle$ and $|L+1-c\rangle$ are in region $C$ and $A$, respectively). We first {split up the total time interval of duration $t$ into $M$ time steps of size $t/M$ and insert a complete set of states after each step~\cite{sakurai}:}
\begin{widetext}
\begin{eqnarray}
    \langle c |e^{-iH t}|a\rangle &=& \sum_{s_1=1}^L \cdots \!\!\! \sum_{s_{M-1}=1}^L 
    \langle c|e^{-iH t/M}|s_{M-1} \rangle  \langle s_{M-1}|e^{-iH t/M}|s_{M-2}\rangle  \cdots \langle s_2|e^{-iH t/M}|s_1\rangle   \langle s_1|e^{-iH t/M}|a\rangle \nonumber  \\
    &=& \sum_{s_1=1}^L \cdots \!\!\! \sum_{s_{M-1}=1}^L \;\; \prod_{r=1}^M \langle s_r|e^{-iH t/M}|s_{r-1}\rangle \,,
\end{eqnarray}
where $|s_0\rangle=|a\rangle$ and $|s_M \rangle=|c\rangle$.
{We may rewrite this as a sum over sequences $(a,s_1,s_2,\cdots s_{M-1},c)$:
\begin{equation}
    \langle c |e^{-iH t}|a\rangle =
    \sum_{(s_0=a,s_1,s_2,\cdots s_{M-1},s_M=c):\; 1 \le s_j  \le L}\;\; \prod_{r=1}^M \langle s_r|e^{-iH t/M}|s_{r-1}\rangle \,. 
\end{equation}
Each sequence corresponds to a possible evolution for the system from site $a$ at time $0$ to site $c$ at time $t$ while going through sites $s_1, s_2,\ldots$ at intermediate times $t/M,2t/M,\ldots$.}
Now for large $M$ we expand as in Euler's method,
\begin{equation} 
\langle c |e^{-iH t}|a\rangle = \sum_{(s_0=a,s_1,s_2,\cdots s_{M-1},s_M=c):\; 1 \le s_j  \le L} \;\;\prod_{r=1}^M \langle s_r|1-iH t/M|s_{r-1}\rangle +O(1/M) \,,
\label{sumpaths}
\end{equation}
\end{widetext}
and since $H$ includes only nearest-neighbor hopping, {$\langle s_r|1-iH t/M|s_{r-1}\rangle$ vanishes for $|s_r-s_{r-1}|>1$. Thus,
the only sequences $(s_0,s_1,s_2,\cdots s_{M-1},s_M)$ that contribute to the sum in the large-$M$ limit are those} where $|s_{r+1}-s_r| \le 1$.\

Consider one such {sequence} $P{=}(s_0{=}a,s_1, s_2, \cdots \!,s_{M-1}, s_M{=}c)$ contributing to the sum in Supplementary Equation~\eqref{sumpaths}.  There will be some time step $R$ at which the {sequence} $P$ first enters region $B$ (and thus $s_R$ is the leftmost site in region $B$ while $s_{R-1}$ is the rightmost site in region $A$). Similarly, let $R'$ be the last time step at which the {sequence} $P$ is in region $B$ (so $s_{R'}$ is the rightmost site in region $B$ while $s_{R'+1}$ is the leftmost site in region $C$). {We may write out the sequence $P$ in more detail as $P=(s_0, s_1, \cdots, s_{R-2},s_{R-1}, s_{R},$} {$s_{R+1}, \cdots s_{R'-2}, s_{R'-1}, s_{R'}, s_{R'+1},s_{R'+2}, \cdots ,s_{M-1},s_{M})$.}

Now we construct a twin {sequence} $\tilde P$ that starts from site $L+1-a$ in region $C$ and ends at site $L+1-c$ in region $A$:
$\tilde P=(L+1-s_0, L+1-s_1, \cdots, L+1-s_{R-2},L+1-s_{R-1}, s_{R'}, s_{R'-1}, s_{R'-2}, \cdots, s_{R+1},s_R, L+1-s_{R'+1},L+1-s_{R'+2}, \cdots ,L+1-s_{M-1},L+1-s_{M})$. The first portion of {sequence} $\tilde P$ is the mirror image (in region $C$) of the first portion of {sequence} $P$ (in region $A$), the middle portion of {sequence} $\tilde P$ (which includes all visits to region $B$) is just the middle portion of {sequence} $P$ traversed in reverse order, and the last portion of {sequence} $\tilde P$ (in region $A$) is the mirror image of the last portion of {sequence} $P$ (in region $C$). {Now sequences $P$ and $\tilde P$ visit the same sites in region $B$ (in reverse order), and $H$ is real symmetric, so reversing the order in which sites are visited does not change the amplitude, $\langle s_r|1-iH t/M|s_{r-1}\rangle=
\langle s_{r-1}|1-iH t/M|s_{r}\rangle$. Thus,}
the amplitude associated with {sequence} $\tilde P$ is identical to the amplitude associated with {sequence} $P$. Since each {sequence} $P$ contributing to $\langle c |e^{-iH t}|a\rangle$ in Supplementary Equation~(\ref{sumpaths}) has a twin $\tilde P$ contributing to $\langle L+1-c |e^{-iH t}|L+1-a\rangle$, {taking $M \to \infty$} we have 
\begin{equation}
\langle c |e^{-iH t}|a\rangle=\langle L+1-c |e^{-iH t}|L+1-a\rangle \,.
\label{sym11}
    \end{equation}

Now we may generalize Supplementary Equation~\eqref{sym11} to a superposition initial state 
$\ket{\Psi(0)}=\sum_{a=1}^{L_A} \kappa_a |a\rangle$
spread over several sites in region $A$. By
linearity,
\begin{equation}
\langle c |e^{-iH t}\ket{\Psi(0)}=\langle L+1-c |e^{-iH t}\ket{\widetilde{\Psi}(0)} \,,
    \end{equation}
where $\ket{\widetilde{\Psi}(0)}= \sum_{a=1}^{L_A} \kappa_{a} |L+1-a\rangle$ is the mirror image under parity of state $\ket{\Psi(0)}$. 

Finally, we may sum over the final state,
\begin{equation}
\sum_{c \in C} \left|\langle c |e^{-iH t}\ket{\Psi(0)}\right|^2=
\sum_{c' \in A} \left|\langle c' |e^{-iH t}\ket{\widetilde{\Psi}(0)}\right|^2 \,,
\end{equation}
proving the theorem.

\end{proof}

\suppnote{3}{Time needed to trap a particle}

\begin{figure}
\begin{center}    
\includegraphics[width=1\linewidth]{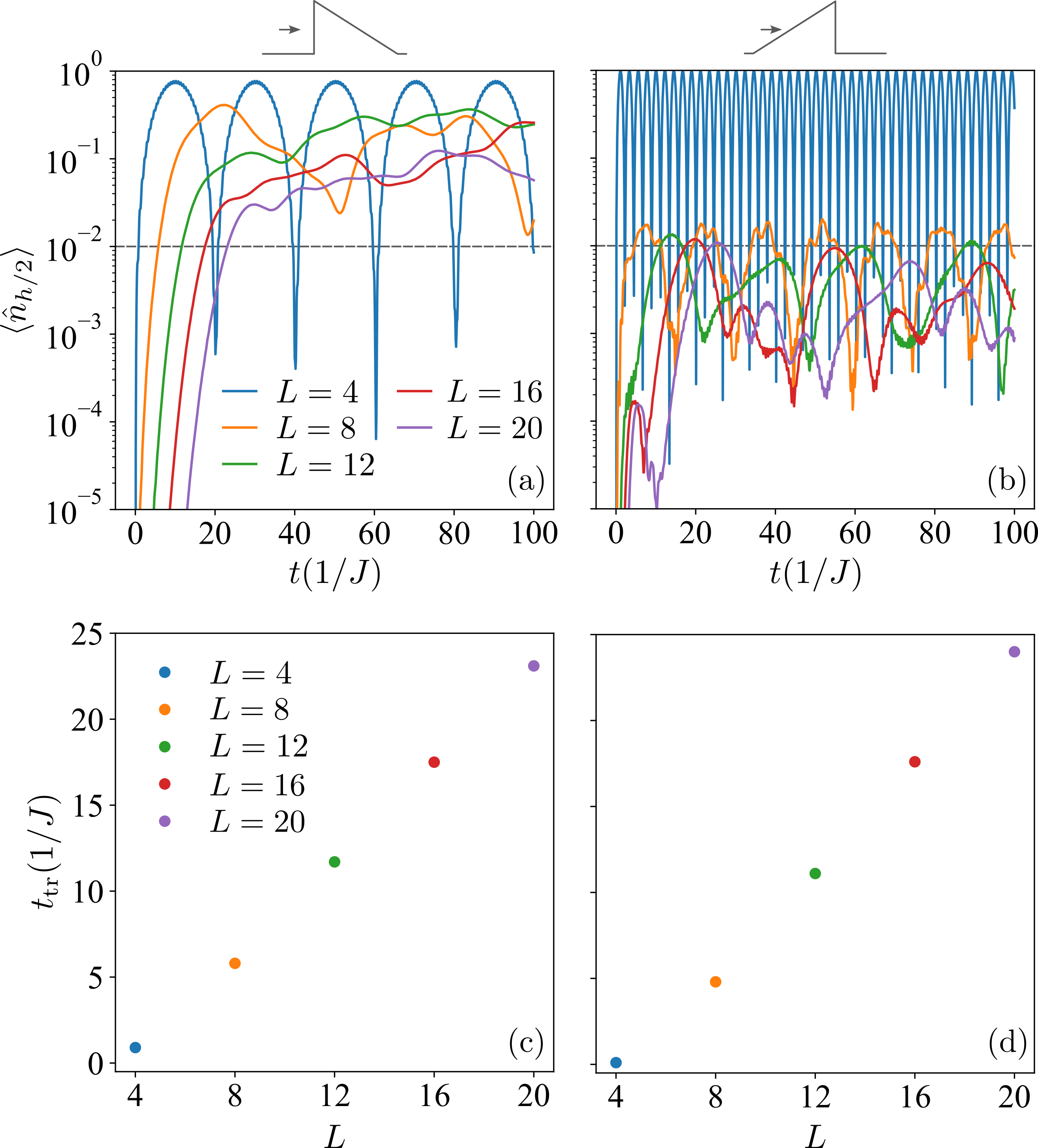}
    \caption{\textbf{Scaling of the time to reach site $j^\ast$ within underbarrier trapping regime.} (a) and (b): Time dependence of $\langle \hat{n}_{h/2} \rangle$ for different system sizes. Dashed gray lines correspond to $\langle \hat{n}_{h/2} \rangle = 0.01$. (c) and (d): Time $t_{\rm tr}$ needed to reach site $j^\ast$ as a function of $L$, i.e., moment of time when $\langle \hat{n}_{h/2} \rangle$ exceeds $0.01$. The right (left) column corresponds to the vertical (angled) orientation of the barrier. For numerical simulations, we used the following parameters: $U=10J$, $h = 20J$. The initial configuration contains a doublon at the leftmost site of the system. }
    \label{fig:tscale_jstar}
\end{center}
\end{figure}

Let us consider the regime of the underbarrier resonant trapping ($U=h/2$) with the initial state
\begin{equation}
    \ket{\Psi(0)}=\hat{c}_{1, \downarrow}^\dagger\hat{c}_{1, \uparrow}^\dagger|0\rangle,
\end{equation}
i.e., a doublon positioned at the leftmost site of the lattice of any size. We define $j^\ast$  as a site where the height of the barrier satisfies $V^{\rm ex}_{j^\ast}=h/2$. Supplementary Figures~\ref{fig:tscale_jstar}~(a) and~(b) show the time evolution of $\langle \hat{n}_{h/2} \rangle$. As the system size increases, the particles have more options to distribute before reaching the barrier, resulting in lower values of $\langle \hat{n}_{h/2} \rangle$. 
Additionaly, for larger system sizes, it takes longer for a particle to reach the site $j^\ast$. Supplementary Figures~\ref{fig:tscale_jstar}~(c) and~(d) illustrate how the time $t_{\rm tr}$ required to trap a particle at $j^\ast$ scales with the system size $L$, where $t_{\rm tr}$ is identified as the moment when $\langle \hat{n}_{h/2} \rangle$ first exceeds 0.01. As can be seen, it displays nearly linear dependence on $L$, except for $L = 4$, where the deviation is attributed to finite-size effects.

    \onecolumngrid
    \section*{Supplementary References}
    \putbib[supp]
\end{bibunit}

\end{document}